\documentclass[11pt]{JHEP3}
\usepackage{amsfonts,ulem,bbold}   
\usepackage{amssymb,amsmath}%,graphicx,wrapfig,epsfig} 
\newcommand{\be}{\begin{equation}}   
\newcommand{\ee}{\end{equation}}   
\newcommand{\bea}{\begin{eqnarray}}   
\newcommand{\eea}{\end{eqnarray}}   
\newcommand{\p}{\partial}   
\newcommand{\nn}{\nonumber}   
   
\newcommand{\mh}{\mathfrak{h}}   
\newcommand{\mH}{\mathbb H}   
\newcommand{\mK}{\mathbb K}   
\newcommand{\mX}{\mathbb X}   
\newcommand{\one}{\mathbb 1}   
   
\DeclareMathOperator{\Tr}{Tr}   

\def\half{\frac{1}{2}}   
\title{On Non-Linear Actions for Massive Gravity}  
   
\author{S. F. Hassan\\ Department of Physics \& The Oskar  
Klein Centre,\\ Stockholm University, AlbaNova University Centre,  
SE-106 91 Stockholm, Sweden \\ E-mail: \email{fawad@fysik.su.se}}  
  
\author{Rachel A. Rosen\\ Department of Physics \& The Oskar Klein  
  Centre,\\ Stockholm University, AlbaNova University Centre, SE-106  
  91 Stockholm, Sweden \\ E-mail: \email{rarosen@fysik.su.se}}  
  
\abstract{   
In this work we present a systematic construction of the potentially 
ghost-free non-linear massive gravity actions. The most general action 
can be regarded as a 2-parameter deformation of a minimal massive 
action. Further extensions vanish in 4 dimensions. The general mass 
term is constructed in terms of a ``deformed'' determinant from which 
this property can clearly be seen. In addition, our formulation 
identifies non-dynamical terms that appear in previous constructions 
and which do not contribute to the equations of motion. We elaborate 
on the formal structure of these theories as well as some of their 
implications. 
}   
\keywords{massive gravity}  
   
\preprint{}   
 
\begin{document}    
   
\section{Introduction and Summary of Results}   
   
There are several motivations for studying massive gravity. From the   
observational point of view, the conclusion that $95\%$ of the   
universe is made of dark energy and dark matter is closely related to   
the assumption that general relativity is equally valid at all length   
scales. It is possible that modifications of general relativity over   
large distances change this picture and massive gravity is one such   
modification. It is also possible that a graviton mass ameliorates the   
cosmological constant problem. From the theoretical perspective, the   
shear difficulty of constructing a consistent theory of massive   
gravity makes the problem more interesting. 
   
In $1939$, Fierz and Pauli \cite{FP1,FP2} constructed the ghost-free   
theory of non-interacting massive gravitons in a flat background.   
Since then, this theory has stubbornly resisted a non-linear   
generalization. The work of Boulware and Deser \cite{BD} showed that a   
generic extension of the Fierz-Pauli (FP) theory to curved backgrounds   
will contain ghost instabilities. The ghost analysis in massive   
gravity was greatly simplified by the work of \cite{AGS}, motivated by   
the Goldstone-vector boson equivalence theorem in quantum field   
theory. The authors in \cite{AGS} argued that, since a graviton mass   
can be associated with a spontaneous breaking of general covariance,   
the associated Goldstone modes in a non-unitary gauge will contain the   
ghost information. Based on this observation, a procedure was outlined   
in \cite{AGS,CNPT} where the ghost could be avoided order-by-order   
by tuning the coefficients in an expansion of the mass term in powers   
of the metric perturbation $h_{\mu\nu}$ and the longitudinal Goldstone   
mode $\pi$. An added advantage is that within this effective field   
theory-like framework, if a ghost turns out to be heavier than the UV   
cutoff of the effective theory it can be safely ignored. Thus even   
when the ghost is not completely avoidable, it could still be   
manageable. The preliminary conclusion in \cite{CNPT} was that at the   
quartic order in Goldstone modes, the expansion will unavoidably   
contain ghosts.   
   
This problem was recently revisited by de Rham and Gabadadze
\cite{dRG2} who showed that massive gravity ghosts can indeed be   
avoided, or at least their appearance relegated to higher energy   
scales where the effective field theory may no longer be applicable.   
One can then successfully construct a theory that is ghost-free in the   
so-called decoupling limit, which amounts to retaining terms only   
first order in $h_{\mu\nu}$, but all orders in $\pi$ \cite{dRG2,dRGT}.   
In particular, \cite{dRGT} constructed a non-linear massive action   
starting from the decoupling limit data. Using this theory it becomes   
possible to study the ghost away from the decoupling limit. It was   
concluded that one of the resummed theories is ghost-free at least up to  
quartic order in the nonlinearities.  
   
In this work we present a systematic construction of massive gravity 
actions that simplifies the analysis and brings out several important 
features. The main results are summarized below. The implications of 
these results are discussed in more detail in the concluding section. 
\begin{itemize}   
\item We construct the most general, potentially ghost-free, mass term 
  as a ``deformed'' determinant. The general massive gravity action 
  can be regarded as a $2$-parameter deformation of a {\it minimal} 
  massive action which is, in turn, the minimal non-linear extension 
  of Fierz-Pauli massive gravity. The origin of the $2$ parameters can 
  be easily seen in this formulation. No further generalizations exist 
  in $4$ dimensions. The full nonlinear equations of motion are also 
  written down.  
\item It is shown that the earlier constructions of non-linear massive
  actions lead to non-dynamical terms in the action that do not
  contribute to the equations of motion. These terms arise due to an
  assumption made to simplify the construction. In our formulation it
  becomes straightforward to isolate and remove these terms, implying
  that the resummed theory found in \cite{dRGT} has an equivalent
  description in terms of a simpler action.  This also shows that
  while the massive action was obtained starting with a covariant
  version of the Fierz-Pauli mass, it also turns out to describe
  theories which reduce to the non-covariant Fierz-Pauli mass at
  lowest order and, hence, is more general than it seems.
\item We emphasize the physical relevance of generalizing the auxiliary
  metric $\eta_{\mu\nu}$, used in the construction, to a general
  $f_{\mu\nu}$. In this context, we discuss solutions of the minimal
  theory that have the potential to avoid the observational
  consequences of the vDVZ discontinuity \cite{vDVZ1,vDVZ2} as well as
  screen a large cosmological constant. 
\end{itemize}    
   
The actions presented here are known to be ghost-free at the
interactive level to first order in the metric perturbation
$h_{\mu\nu}$ and to all orders in the Goldstone longitudinal mode,
$\pi$. This was extended to quartic order by an ADM analysis in
\cite{dRGT}. However, a proof that these actions are ghost free at the
complete non-linear level is still lacking (see, for example,
\cite{CM3} for an argument that the ghost is in fact present at the
quartic order). But the stronger statement is that if any such
non-linear ghost free action exists, it will have to belong to this
family of actions (see note added at the end).
   
The organization of this paper is as follows.  In section 2 we review
the basic features of massive gravity, paying particular attention to
the features that will be most relevant when constructing a nonlinear
theory. We also briefly summarize the recent developments in
constructing ghost-free nonlinear extensions of the FP action. In
section 3 we construct the minimal non-linear extension of the
Fierz-Pauli mass term and discuss its screening solutions as well as
the origin of non-dynamical terms. In section 4 the generalization of
this model is considered.  We describe the formal structure of the
theory and argue that there exists at most a $2$-parameter family of
actions with the desired properties.  We write down the full
non-linear equations of motion for the generalized theory. Some
implications of our results are discussed in section 5.

\section{General Structure of non-Linear Massive Gravity}   
   
In this section we review features of massive gravity that are   
relevant to the construction of non-linear mass terms, and make some   
observations that clarify the underlying structures. We then briefly   
discuss recent developments in constructing ghost-free non-linear   
extensions of Fierz-Pauli massive gravity.  This provides the   
background and conventions for the systematic approach of the   
following sections.   
   
\subsection{The Fierz-Pauli mass in linear theory}   
Fierz and Pauli \cite{FP1,FP2} constructed a mass term for linearized  
metric fluctuations around flat spacetime,   
$h_{\mu\nu}=g_{\mu\nu}-\eta_{\mu\nu}$,    
\be   
\frac{m^2}{4}\,\left(h_{\mu\nu}h^{\mu\nu}-a\, h^\mu_\mu h^\nu_\nu\right)\,,   
\qquad {\rm with} \quad a=1\,.  
\label{FP}   
\ee    
For any other value of $a$ the theory has a ghost.  Invariance under  
infinitesimal general coordinate transformations (GCT) can be restored  
by introducing St\"uckelberg fields and replacing $h_{\mu\nu}$ by  
$h_{\mu\nu}+\p_\mu\pi_\nu+\p_\nu\pi_\mu$.  In the St\"uckelberg  
formulation, the longitudinal mode $\pi$ defined through  
$\pi_\mu=\pi_\mu^\perp+\p_\mu \pi$, appears in the action through a  
$4$-derivative term and carries the ghost degree of freedom.  The  
ghost decouples only for $a=1$. Unfortunately, unlike the case of  
massless linearized general relativity, there is no obvious unique way  
to extend the massive linearized theory to the non-linear level. But  
in recent work, ghost decoupling has turned out to be a powerful  
criterion. Let us first discuss the general features of a non-linear  
mass term before discussing the recent developments towards its  
construction.  
   
\subsection{The auxiliary metric and the inverse metric}   
   
On general grounds, a non-linear mass term in gravity (a) cannot   
contain derivatives of $g_{\mu\nu}$ and (b) should be invariant under   
general coordinate transformations (GCT). Such a term cannot be   
constructed using the metric alone since the only two relevant   
quantities, $\det g$ and ${\rm Tr} g=4$, are not adequate for this   
purpose \cite{BD}. To overcome this, one of the following options may   
be used:   
\begin{enumerate}   
\item Introduce an extra rank two tensor $f_{\mu\nu}$ as in   
  \cite{BD,AGS}.   
\item Introduce an auxiliary extra dimension \cite{G,dR}.   
\item Generate a mass through a gravitational Higgs
  mechanism \footnote{Note that in this context, the term Higgs
    mechanism does not imply the existence of a mass generation
    mechanism on par with that in vector gauge theories, and the
    corresponding ramifications for unitarity, {\it etc.} Here, it
    mainly refers to the association of the graviton mass with scalars
    of the form $\phi^\mu=x^\mu+\delta\phi$. We retain the terminology
    with this caveat.}\cite{GT,S,tHooft,Kakushadze,CM1,CM2}.
\end{enumerate}   
Both options (2) and (3) can be formulated in terms of (1) as will be   
discussed below for Higgs gravity and in \cite{HR} for the extra   
auxiliary dimension model. Hence we will concentrate on this first   
case.    
   
Combining the dynamical metric $g_{\mu\nu}$ with a covariant rank two   
tensor $f_{\mu\nu}(x)$ (henceforth called the auxiliary metric), one   
can construct arbitrary scalar functions $F(g^{-1}f)$ and   
$F'(f^{-1}g)$. The GCT invariant mass term has to be written in terms   
such functions.  A special case is when the mass term involves only   
$F(g^{-1}f)$, {\it i.e.,} it depends only on the {\it inverse metric}   
$g^{\mu\nu}$. A simple inspection shows that the massive gravity   
actions considered in \cite{AGS}, and in the subsequent works   
\cite{CNPT,dRG2,dRGT}, all belong to this class \footnote{This also   
  holds for the ``auxiliary extra dimension'' model of   
  \cite{G,dR,dRG1} which is special in that it admits a description in   
  terms of both $g^{-1}f$ and $f^{-1}g$ \cite{HR}.}, even though,   
in practice, the setups may have been described in terms of   
$g_{\mu\nu}$. In the last paragraph of this section we argue that the   
combination $g^{-1}f$ is adequate to describe any ghost free mass term   
and so will be explicitly assumed here. Thus we consider massive   
gravity actions of the form, \be S=-M_p^2\int d^4x\sqrt{-g}\,R(g) +   
M_p^2 m^2\int d^4x \sqrt{-g}\,F(g^{-1} f)\,.   
\label{F}   
\ee       
Invariance under general coordinate transformations implies that $F$   
must be expandable as a sum of products of the type ${\rm   
  Tr}[(g^{-1}f)^n]$.   
   
Invariance under GCT is insured since $f_{\mu\nu}$ transforms as a   
rank 2 tensor. To make this explicit, one can parametrize the   
auxiliary metric as a coordinate transformation of some fixed metric   
\cite{AGS},   
\be   
f_{\mu\nu}(x)=\frac{\p \phi^a}{\p x^\mu}\,\bar   
f_{ab}(\phi)\, \frac{\p \phi^b}{\p x^\nu}\,.   
\label{fphi}   
\ee    
The $\phi^a(x)$ are to be regarded as general coordinate   
transformations of the $x^\mu$ and generalize St\"uckelberg fields to   
non-linear level. $\bar f_{ab}$ is a fixed auxiliary metric that   
should be chosen consistent with the physics of the problem. Both   
these are scalar functions \footnote{To see this, start with a   
  coordinate system $\{\phi^\mu\}$ and consider two   
  reparametrizations $\{x^\mu\}$ and $\{x'^\mu\}$ defined by   
  $\phi^\mu(x)$ and $\phi'^\mu(x')$. Now, a coordinate transformation   
  from $x$ to $x'$ is subject to $\phi^\mu(x)=\phi'^\mu(x')$   
  clarifying that $\phi^\mu(x)$ are scalars.} of $x^\mu$ ($\bar f$ is    
a tensor with respect to reparametrizations of $\phi^a$).  The   
physical (unitary) gauge corresponds to choosing $\phi^a(x)=x^a$.  In   
\cite{BD} $\bar f$ is identified with a background dynamical metric   
$\bar g$ in the physical gauge. If $\bar f$ is taken to be a   
flat metric, the $\phi^a$ can be chosen such that $\bar   
f_{ab}=\eta_{ab}$.   
   
\subsection{Non-linear generalizations of the Fierz-Pauli   
  mass}    
   
Let us now describe the ideas that have recently been used to   
construct non-linear graviton mass terms \footnote{While the   
  structures introduced in this subsection are useful for motivating   
  the form of the non-linear actions of later sections, and for seeing   
  the relation with other recent work, they are not needed for   
  formulating the non-linear theory, which is good news in view of   
  footnote \ref{note2}}. Given the auxiliary metric $f_{\mu\nu}$, a   
covariant non-linear version of the Fierz-Pauli mass (\ref{FP}) can be   
constructed in terms of a matrix $H^\mu_\nu$ given by,    
\be   
g^{\mu\lambda}f_{\lambda\nu}=\delta^\mu_\nu - H^\mu_\nu\,.   
\label{H}   
\ee    
This is obviously a $(1,1)$ tensor under GCT. Note that, rewritten as    
$H_{\mu\nu}\equiv g_{\mu\lambda}H^\lambda_\nu =g_{\mu\nu}-f_{\mu\nu}$,   
one recovers the matrix introduced in \cite{AGS} in a more elaborate   
setup. Any mass term can be expressed entirely in terms of   
$H^\mu_\nu$, and we will only retain this in the description. Then the   
Fierz-Pauli mass can be generalized to a covariant non-linear form   
(see \cite{AGS} and references therein),      
\be   
%m^2\int d^4x\sqrt{-g}\,(H^\mu_\nu H^\nu_\mu-H^\mu_\mu H^\nu_\nu)\,,   
\frac{m^2}{4}\,\sqrt{-g}\,\left(H^\mu_\nu H^\nu_\mu-H^\mu_\mu H^\nu_\nu  
\right)\,,    
\label{FPCov}   
\ee   
which reduces to the original FP mass (\ref{FP}) at the quadratic   
order in $h_{\mu\nu}$ and in the unitary gauge for $\phi^a$. We refer   
to this as the {\it covariant} FP term.     
   
But this is not enough. Though (\ref{FPCov}) is ghost free at the   
quadratic order around flat background, the ghost mode generally   
reappears at the quadratic order around curved backgrounds   
\cite{BD}. Equivalently stated, the ghost reappears at higher orders   
around flat background.  To cure this problem, the idea, as proposed   
in \cite{AGS} and pursued in \cite{CNPT,dRG2}, is to add terms to   
(\ref{FPCov}) that are higher order in $H^\mu_\nu$ to cancel the ghost   
at higher orders.   
   
To simplify the ghost analysis, the theory is often written as an   
expansion around flat background, which for $g^{\mu\nu}$ gives,   
\be   
g^{\mu\nu}=\eta^{\mu\nu}+{\mathfrak h}^{\mu\nu}\,.   
\ee    
In terms of $h_{\mu\nu}=g_{\mu\nu}-\eta_{\mu\nu}$, one has   
${\mathfrak h}^{\mu\nu}=\sum_{n=1}^\infty \left[(-{\eta}^{\,-1}h)^n    
\eta^{\,-1}\right]^{\mu\nu}$. Also, since the $\phi^a(x)$ are general    
coordinate transformations of $x^\mu$, they can be expanded around the    
``unit transformation'' as \cite{AGS},   
\be   
\phi^\mu(x)=x^\mu+\pi^\mu(x)\,.   
\ee    
The $\pi^\mu$ transform as ``coordinate differences'' under GCT. Then   
$H^\mu_\nu$ given by (\ref{H}) take the form (on choosing $\bar   
f_{\mu\nu}=\eta_{\mu\nu}$ to facilitate comparison with FP theory),       
\be   
H^\mu_\nu=-\p^\mu\pi_\nu-\p_\nu\pi^\mu-\p^\mu\pi_a   
\p_\nu\pi^a - \mh^\mu_\rho\,(\delta^\rho_\nu + \p_\nu\pi^\rho+   
\p^\rho\pi_\nu+\p^\rho\pi_a\p_\nu\pi^a)   
\ee   
Here, all indices have been raised or lowered with respect to their   
natural positions with the background metric $\eta_{\mu\nu}$. The   
expression is exact in $\pi^\mu$ and in $\mh^{\mu\nu}$ which, so far,   
need not be infinitesimal. If the graviton mass is assumed to arise as   
a result of a spontaneous breaking of GCT, through some kind of   
gravitational Higgs mechanism, then the $\pi^a$ would be the unphysical   
(pure gauge) Goldstone modes associated with the symmetry breaking 
\cite{GT,S}.    
   
The field most relevant to the ghost problem is the ``longitudinal''   
mode $\pi$, given by,     
$$   
\pi_\mu=\pi^\perp_\mu+\p_\mu\pi\,.   
$$   
Let's use the notation $\mathbb H$ for the matrix with elements   
$H^\mu_\nu$ to avoid mixup with $H=H^\mu_\mu$. Also, let $\Pi$   
denote the matrix with elements $\Pi^\mu_\nu=\p^\mu\p_\nu\pi$. Then   
retaining only $\pi$, and setting $\pi_\mu^\perp=0$, one has,     
\be   
g^{-1}f\equiv (\mathbb 1-\mathbb H)=   
(\mathbb 1+\mh\eta) (\mathbb 1+\Pi)^2\,.   
\label{gfPi}   
\ee  
This expression is exact \footnote{A word caution on covariance 
  issues: The above parametrization of fluctuations refers to a 
  perturbative expansion around flat background. At the non-linear 
  level the $\pi^\mu$ are {\it coordinate differences} and hence 
  $\pi_\mu^\perp$ and $\pi$ are not proper vectors and scalars. Thus, 
  strictly speaking, statements that hold for vector and scalar modes 
  at the linearized level, may not fully apply here. Also, the 
  resulting expressions cannot be covariantized by simply using 
  covariant derivatives. If one insists on using this type of 
  parametrization for the non-linear theory, while maintaining 
  manifest general covariance, then an option is to use Riemann normal 
  coordinates and express $\pi^\mu$ in terms of a geodesic tangent 
  vector $u^\mu$ and derivatives of the curvature tensor as 
  in\cite{Alvarez-Gaume-Freedman-Mukhi}.\label{note2}} to all orders 
in $\mh$ and $\pi$. 
   
{\bf Recent developments:} As discussed in \cite{AGS,CNPT}, the  
Boulware-Deser ghost instability of massive gravity is associated with  
the appearance of $\pi$ dependent terms in the action with more than  
$2$ time derivatives. Thus, in the massive gravity action (\ref{F}),  
an expansion of $F(\mathbb 1-\mathbb H)$ in powers of $\mH$ should be  
constructed in such a way as to avoid this problem. Such a program was  
carried out successfully by de Rham and Gabadadze  in \cite{dRG2}.  
They obtained a $2$ parameter family of actions to quintic order in  
$\mH$ that avoids the ghost problem at least up to first order in  
$h_{\mu\nu}$ and all orders in $\pi$. Subsequently \cite{dRGT} 
proposed a method for resumming such expansions in powers of $\mH$ 
into closed-form non-linear actions.   
   
In the next two sections, we present an alternative formulation of the   
massive action based on a ``deformed determinant'' that clearly brings   
out the systematics, showing in particular that the framework is more   
general than might appear at first sight.     
   
{\bf Adequacy of the description in terms of $g^{-1}f$:} Consider a   
mass term that also depends on $f^{-1}g\equiv (\mathbb 1-\mH)^{-1}$.   
It is obvious that any scalar function $F'(f^{-1}g)$ can also be   
expanded in powers of $\mH$. The quintic order polynomials of $\mH$   
analyzed in \cite{dRG2} are general enough that they do not exclude   
$F'$ in the mass term. However, as can be seem from this work as well   
as from \cite{dRGT}, all ghost free combinations are obtainable from   
actions that contain $F(g^{-1}f)$ alone, owing to the simple relation  
between $\Pi$ and $\mH$ in (\ref{gfPi}). Hence this structure is   
adequate for describing ghost free massive gravity actions. Though,   
up to a given order, the $\mH$-expansions may also be obtainable from   
an $F'(f^{-1}g)$, that would correspond to a complicated rewriting of   
$F(g^{-1}f)$ to that order.   
   
{\bf Relation to Higgs gravity:} In Higgs gravity setups \cite{tHooft,   
  Kakushadze,CM1,CM2}, one considers gravity coupled to $4$ scalar fields,   
say $\phi^a$ with a Minkowskian internal metric. The scalars are   
supposed to acquire a vacuum expectation value $<\phi^a>=x^a$ through   
some dynamics and break the GCT symmetry of the action, rendering the   
gravitons massive \footnote{A dynamical realization of Higgs gravity   
  is provided by the ``brane induced gravity'' construction, as   
  discussed in \cite{HHvS}. While the obvious breaking of translation   
  invariance in directions transverse to the brane generates massless   
  Goldstone modes, a more subtle spontaneous breaking of GCT along the 
  brane occurs as a result of integrating out bulk modes. This     
  automatically generates the $\phi^a$ and the graviton mass in a   
  gauge invariant set up, though a ghost free construction is still   
  lacking.}. Note that all terms involving derivatives of the $\phi^a$   
can be written in terms of $g^{-1}f$, or using the trace structure, in   
terms of $\p_\mu\phi^a g^{\mu\nu}\p_\nu\phi^b\bar f_{bc}$. For   
example, the basic kinetic term is $\Tr(g^{-1}f)$. This accounts for   
the practical similarity between Higgs gravity and massive gravity, as,   
for example, in \cite{BM}.   
   
\section{The Minimal non-Linear Massive Gravity Action}   
   
In this section we construct a minimal non-linear extension of the   
Fierz-Pauli mass term and discuss some of its main features. This is   
useful for understanding the more general non-linear extension to be   
given in the next section.   
   
\subsection{The mass term as a minimally deformed determinant}   
   
The construction of the non-linear graviton mass initiated in   
\cite{AGS,CNPT} and successfully employed in \cite{dRG2,dRGT}, is    
based on two simple criteria.    
\begin{enumerate}   
\item The absence of ghosts at low orders: The function $F(g^{-1}f)$   
  in the massive gravity action (\ref{F}) is constrained by the   
  requirement that its expansion around a flat background leads to a   
  ghost-free theory. The expansion involves $\mh^{\mu\nu}$ and   
  $\pi_\mu=\p_\mu\pi$ (ignoring the ``vector'' modes) through   
  (\ref{gfPi}). As explained in \cite{CNPT}, the ghost is associated   
  with $\pi$ terms with higher than two derivatives which, therefore,   
  should not arise in the expansion of $F$. In particular, the   
  $\mh$-independent part, $F(g^{-1}f)\vert_{h=0}$, should contain the   
  potentially dangerous $\pi$ terms only through total   
  derivatives. The next step is to check that the term linear in $\mh$   
  is also ghost free (this is, of course, necessary but not sufficient   
  for avoiding the ghost to higher orders in $\mh$).   
\item The recovery of the {\it covariant} Fierz-Pauli action   
  (\ref{FPCov}) at the lowest order.   
\end{enumerate}   
Note that the second criterion demands more than the minimum  
requirement of reproducing the original FP mass in flat background  
(\ref{FP}). Hence it could, potentially, exclude non-linear actions  
that reduce to (\ref{FP}) without producing (\ref{FPCov}). We will  
show that these two classes of actions only differ by non-dynamical  
terms that can be easily isolated.  
  
In \cite{CNPT,dRG2,dRGT}, to implement the first criterion, total   
derivative functions of $\Pi^\mu_\nu=\p^\mu\p_\nu\pi$ are explicitly   
constructed in terms of traces of powers of $\Pi$ (\ref{gfPi}). In   
contrast, the approach here is based on the observation that all such   
total derivative terms naturally appear in the expansion of   
$\det(\mathbb 1+\Pi)$. This simplifies both the procedure and the   
final results by identifying the redundancies that result from the   
second criterion. 
   
To satisfy the first criterion, consider $F(g^{-1}f)$ at lowest   
order, {\it i.e.,} for $\mh=0$, when,   
\be   
(g^{-1}f)\big\vert_{h=0}= (\mathbb 1+\Pi)^2.   
\ee   
As mentioned above, at this order $\pi$ must appear in the Lagrangian   
only in total derivative terms.     
%There are only four such terms; total derivative terms fifth order and   
%higher in $\pi$ vanish identically \cite{CNPT}.     
All such total derivative terms appear automatically (with given   
coefficients) in an expansion of the determinant,   
\be   
\det(\mathbb 1+\Pi)= \sum_{n=0}^{4}\frac{-1}{n ! (4-n)!}\,   
\epsilon_{\mu_1\cdots\mu_n\lambda_{n+1}\cdots\lambda_4}\,   
\epsilon^{\nu_1\cdots\nu_n\lambda_{n+1}\cdots\lambda_4}\p^{\mu_1}\p_{\nu_1}\pi   
\cdots \p^{\mu_n}\p_{\nu_n}\pi\,.   
\label{detTD}   
\ee    
The total derivative nature is manifest due to the antisymmetry   
property of the epsilon tensor.  We can choose $F\big\vert_{h=0}$ to    
have this form, but the first two terms of the summation,     
$1+\p^\mu\p_\mu\pi$, (for $n=0$ and $1$, respectively) cannot arise in   
an action that, by the second criterion, reduces to the {\it   
  covariant} FP form (\ref{FPCov}) at the lowest order (since this is   
at least quadratic in $\pi$). Thus choosing $F\vert_{h=0}$   
consistent with the second criterion amounts to deforming   
$\det(\mathbb 1+\Pi)$ by removing the first two terms from the above    
expansion,     
\be   
F(g^{-1}f)\vert_{h=0}\sim \det(\mathbb 1+\Pi)-\Tr(\mathbb 1+\Pi)+3\,.   
\ee   
This is the {\it minimal} deformation of the determinant structure.   
One may also consider non-minimal deformations by changing the   
coefficients of the $n>2$ terms in (\ref{detTD}) which, obviously,   
still preserve their total derivative structure. Here we concentrate   
on the minimal deformation, leaving the general case to the next   
section.   
   
Now, the field $\pi$ enters the action only through $g^{-1}f$, in the   
combination (\ref{gfPi}). In particular, $\det(\mathbb 1+\Pi)=   
\sqrt{\det(g^{-1}f)}\big\vert_{h=0}$ and $\Tr(\mathbb 1+\Pi)=\Tr   
\sqrt{g^{-1}f}\big\vert_{h=0}$. Thus, going beyond $\mh=0$   
requires making the substitution $\mathbb 1+\Pi \rightarrow   
\sqrt{g^{-1}f}$ leading to a massive gravity action (\ref{F}), with   
the non-linear mass term    
\be   
m^2\int d^4x \sqrt{-g}\,F   
=2 m^2\int d^4x \sqrt{-g}\,\left[\sqrt{\det(g^{-1} f)}   
-\Tr \sqrt{g^{-1} f} +3\right]\,.    
\label{MGA}   
\ee       
The construction involves the square root matrix defined such that  
\footnote{For the metric, we stick to the conventional notation   
  $\sqrt{- g}=\sqrt{-\det g}$. In all other cases, $\sqrt{E}$  
  represents the square-root matrix and not the determinant.}  
$\sqrt{E}\sqrt{E}=E$. Again, the mass term is given by a {\it minimal}  
deformation of $\det(\sqrt{g^{-1} f)}= \det(\mathbb 1+\sqrt{g^{-1}  
  f}-\mathbb 1)$ that involves removing the first two terms from its  
expansion.  
  
To confirm that our massive action is the non-linear completion of one  
of the ghost free quintic order polynomial actions of \cite{dRG2}, we  
expand it in powers of $\mH$.  To lowest order in $\mH$, one recovers  
the covariant form of the FP mass term (\ref{FPCov}).  To quintic  
order in $\mH$ one obtains the action presented in \cite{dRG2} with  
the particular coefficients\footnote{To correctly match the  
  coefficients at quintic order we note that there is a typo in  
  \cite{dRG2}.  The expression for $f_3$ should be corrected to  
  $f_3=\frac{3}{8}c_3-3d_5 + 20f_7$.} $c_3=\frac{1}{6},  
d_5=-\frac{1}{48}$ ($f_7$ there multiplies terms that vanish in  
4-dimensions as will be discussed in section 4). Generalizations of  
this action are considered in the next section.  
  
The above action can be trimmed further:   
\begin{enumerate}   
\item The term $3\sqrt{-g}$ is needed to cancel out a cosmological   
  constant of order $M_p^2m^2$. If this cancellation is not required,   
  the factor $3$ can be replaced by a free parameter.   
\item The first term simply reduces to $\sqrt{-\det f}$ and does not   
  contribute to the metric equation of motion. As shown in the   
  next subsection, it also does not contribute to the $\phi^a$   
  equations since these are already contained in the metric equation.   
  Thus, being non-dynamical, it can be dropped from the action at the   
  classical level (as long as one does not intend to promote $\bar  
  f_{\mu\nu}$ to a dynamical variable). The origin of this term will  
  be explained below.     
\end{enumerate}   
This leaves us with an action that turns out to be the {\it minimal   
  non-linear extension} of the massive Fierz-Pauli action in the   
presence of a cosmological constant $\Lambda=\Lambda'+6m^2$,   
\be   
S_{min}=-M_p^2\int d^4x \sqrt{-g}\left[     
R + 2m^2\,\Tr \sqrt{g^{-1} f} + \Lambda'\right]\,.   
\label{MGAmin}   
\ee      
Note that although the action is constructed as a minimal deformation   
of a {\it determinant}, the determinant itself becomes   
non-dynamical and only the ``deformation'' stays on. This action is   
also ``minimal'' in the sense of having the simplest structure as   
compared to the generalizations constructed in the next section.    
  
If we now expand the mass term (\ref{MGAmin}) in powers of $\mH$ we  
find, to quadratic order,   
\be   
2m^2\sqrt{g}\,(1-\half\Tr\mH-\frac{1}{8}\Tr(\mH^2)+{\cal O}(\mH^3))\,,   
\label{MGAminH}  
\ee   
which is not of the covariant FP form (\ref{FPCov}). However,  
expanding this in the metric perturbation $h_{\mu \nu}$ around flat  
spacetime does reproduce the basic FP form (\ref{FP}). Hence, this is  
a valid mass term. What has happened is as follows: As stated above,  
the first term $\sqrt{-\det f}$ in (\ref{MGA}) does not contribute to  
the equations of motion. However, this term is needed to reproduce the  
expansion that starts with the {\it covariant} FP form (\ref{FPCov})  
at the quadratic level, as was required by the construction. But this  
requirement can be relaxed since it is totally acceptable that a  
non-linear massive action directly reduces to the basic Fierz-Pauli  
form (\ref{FP}) at the quadratic level in $h_{\mu\nu}$ without ever  
reproducing (\ref{FPCov}). Thus, one concludes that while taking the  
covariant FP term (\ref{FPCov}) as the starting point simplifies the  
construction, it does not make it less general. Rather, it only leads  
to redundant non-dynamical terms in the action that are present   
for ``cosmetic'' reasons. The formulation presented here makes this  
manifest and conveniently isolates the redundant terms. To drive home  
the point, note that in a formulation that does not take care of the  
redundancy, the same minimal theory looks much more complicated than  
(\ref{MGAmin}), {\it i.e.,} it is given by (\ref{MGA}) where  
$\sqrt{\det(g^{-1}f)}$ is expanded in traces of powers of the  
square-root matrix $\sqrt{g^{-1}f}$.   
   
Returning to the ghost issue, when expanded in powers of $\mh$ around   
flat background, the lowest order term in (\ref{MGA}) or   
(\ref{MGAmin}) is ghost free by construction. The next term, linear in   
$\mh$ and to all orders in $\pi$, is also healthy having the form (see   
footnote \ref{note1} below),    
\be   
\frac{1}{2}m^2\,   
\mh^{\mu\nu}\left(\p_\mu\p_\nu-\eta_{\mu\nu}\Box\right)   
\pi\,.   
\label{hpi}   
\ee   
In particular no higher powers and derivatives of $\pi$ arise at the   
linear order in $\mh$. This corresponds to one of the models   
constructed in \cite{dRG2} to quintic order in $\mH=\mathbb 1-g^{-1}f$.   
It was pointed out there that, in the absence of non-linear $\pi$   
interactions, the theory will not exhibit the Vainshtein mechanism   
\cite{V,DDGV} which is associated with the strong coupling   
behavior of $\pi$ in the limit $m\rightarrow 0$. More precisely,   
non-linear $\pi$ terms at higher orders in $\mh$ may still generate a   
strong coupling scale but this will be suppressed by the Planck scale,   
leading to a much smaller Vainshtein radius. The issue of ghosts at   
higher orders in $\mh$ is still an open question and will not be   
addressed here.   
   
\subsection{Equations of motion}   
   
In this subsection we write down the $g^{\mu\nu}$ and $\phi^a$   
equations of motion and, in particular, show that the ``cosmetic''     
$\sqrt{-\det f}$ term in (\ref{MGA}) does not contribute to the   
$\phi^a$ equations.    
   
{\bf The metric equation:} This can be worked out using (\ref{MGAmin})   
and reads \footnote{Note that $E=(\sqrt{E})^2$ implies $\delta   
  E=(\delta\sqrt{E}) \sqrt{E}+ \sqrt{E}(\delta\sqrt{E})$, so that    
  $\Tr(\delta\sqrt{E})=(1/2) \Tr[(\sqrt{E})^{-1}\delta E]$. This last   
  equation holds only under the trace. Then for $E=g^{-1}f$, a   
  variation of the metric gives $\Tr(\delta\sqrt{g^{-1}f)}=   
  (1/4)\Tr\left[\left(g\sqrt{g^{-1}f}+   
    (\sqrt{g^{-1}f})^Tg\right)\delta g^{-1}\right]$, where use is made   
  of the cyclic property of the trace and $f(\sqrt{g^{-1}f})^{-1}=g   
  (g^{-1}f)(\sqrt{g^{-1}f})^{-1}=g\sqrt{g^{-1}f}$. The result is then   
  explicitly symmetrized. \label{note1}},   
\bea   
R_{\mu\nu}-\frac{1}{2}g_{\mu\nu}R -\frac{1}{2}\Lambda g_{\mu\nu}   
&+&   
\frac{1}{2}m^2\left[g\sqrt{g^{-1}f}+(\sqrt{g^{-1}f})^Tg\right]_{\mu\nu}    
\nn\\  
&&\quad +\, m^2 g_{\mu\nu} \left(3-\Tr\sqrt{g^{-1}f}\right)   
=G_N T_{\mu\nu} \,.   
\label{eom-g}   
\eea   
At least for a diagonal metric ansatz, as in \cite{BDZ}, the square-root  
matrix simplifies and one gets closed-form non-linear differential  
equations that could  in principle be solved.   
  
The left hand side is not automatically divergence free, rather its  
divergence yields,      
\be   
\nabla_\mu(\sqrt{g^{-1}f})^\mu_{\,\,\nu} + g^{\mu\rho}g_{\nu\sigma}    
\nabla_\mu(\sqrt{g^{-1}f})^\sigma_{\,\,\rho}    
-2\p_\nu  \Tr(\sqrt{g^{-1}f}) =0\,.   
\label{div}   
\ee   
Here the covariant derivatives are to be written out regarding   
$\sqrt{g^{-1}f}$ as a $(1,1)$ tensor, similar to $g^{-1}f$. This   
follows from the expansion of the radical in powers of $\mH$, or   
from the fact that,   
using $E=(\sqrt{E})^2$,   
a   
transformation $E\rightarrow AEA^{-1}$ implies $\sqrt{E}\rightarrow   
A\sqrt{E} A^{-1}$. The trace of (\ref{eom-g}) gives,   
\be   
R+2\Lambda+3m^2\left(\Tr\sqrt{g^{-1}f} -4\right)=G_N T^\mu_\mu\,.   
\label{tr-eom-g}   
\ee   
A simple class of solutions will be considered below.  It is  
instructive to contrast these equations with their linearized forms  
around flat spacetime (the FP case). The linear version of (\ref{div})  
implies the vanishing of the linearized curvature scalar $R_L$. A  
non-zero $R_L$ cannot be recovered in the limit $m^2\rightarrow 0$  
which is at the origin of the vDVZ discontinuity of linearized massive  
gravity.  
   
{\bf The $\phi^a$ equations:} These are equivalent to the divergence of   
the metric equation (\ref{div}). This is a consequence of the general   
covariance of the action $S_{min}[g,\phi]$. Indeed, under a general   
coordinate transformation, $\delta x^\mu=\xi^\mu$, $\delta g^{\mu\nu}=   
-2\nabla^{(\mu}\xi^{\nu)}$ and $\delta\phi^a=-\xi^\mu\p_\mu\phi^a$.   
The invariance of the action then implies $\int   
d^4x\sqrt{-g}[(\frac{2}{\sqrt{-g}}\frac{\delta S} {\delta   
    g^{\mu\nu}})\nabla^\mu\xi^\nu + \frac{1}{\sqrt{-g}}\frac{\delta   
    S}{\delta\phi^a} \p_\nu\phi^a\xi^\nu]=0$, or on integrating by   
parts,   
\be 
\nabla^\mu(\frac{2}{\sqrt{-g}}\frac{\delta S}{\delta g^{\mu\nu}})   
=\frac{1}{\sqrt{-g}}\frac{\delta S}{\delta\phi^a}\p_\nu\phi^a\,.   
\ee 
The left hand side is the divergence of (\ref{eom-g}). Since  
$\phi^a(x)$ are non-singular coordinate transforms of $x^\mu$, the  
matrix $\p_\nu\phi^a$ is invertible. Hence the $\phi^a$ equations  
$\delta S/\delta\phi^a=0$ are equivalent to the divergence equations  
(\ref{div}). As a consequence, the $\phi^a$ equations are not affected  
by $\sqrt{-\det f}$ in (\ref{MGA}), since it does not contribute to  
the metric equation. It is instructive to check this explicitly. The  
$\phi^a$ equations of (\ref{MGA}), including $\sqrt{-\det f}$, are,   
\be   
\nabla_\mu(D^{\mu\nu}\p_\nu\phi^a\bar f_{ac}) + D^{\mu\nu}   
\p_\nu\phi^a\frac{\delta\bar f_{ab}}{\delta\phi^c}\p_\mu\phi^b=0\,,   
\ee   
where,    
\be 
D^{\mu\nu}=\sqrt{\det(g^{-1}f)}(f^{-1})^{\mu\nu}-\frac{1}{2}\left[   
\sqrt{g^{-1}f}+g^{-1}(\sqrt{g^{-1}f})^Tg\right]^\mu_{\,\,\,\,\rho}   
(f^{-1})^{\rho\nu}\,.   
\ee 
Multiplying by the invertible matrix $\p_\lambda\phi^c$ and   
manipulating, this reduces to the equivalent expression (\ref{div}),   
as expected. In particular, the contribution from the first term of   
$D^{\mu\nu}$ completely drops out, once again showing that  
$\sqrt{-\det f}$ is non-dynamical.     
   
\subsection{The screening solutions}   
   
For the massive gravity actions considered here, the proof of absence  
of ghosts at low orders is valid only for a flat auxiliary metric,  
$\bar f_{\mu\nu}=\eta_{\mu\nu}$. This theory exhibits the vDVZ  
discontinuity \cite{vDVZ1,vDVZ2} at the linear order which implies  
radically different bending of light near massive objects as compared  
to general relativity. This difference needs to be avoided at the  
non-linear level, at least in the vicinity of massive objects, should  
a theory of light gravitons describe the real world. In general,  
massive gravity can achieve this via the Vainshtein  
mechanism\footnote{But even in that case, the vDVZ discontinuity can  
  be tested far away from massive sources. For example, by comparing  
  mass estimates for galaxies and galaxy clusters as derived from  
  gravitational lensing, with dynamical mass estimates, such models  
  can be constrained observationally \cite{Sjors}.}  
\cite{V,DDGV}. However, the analysis of \cite{dRG2}, performed in the  
so-called {\it decoupling limit} approximation, indicates that this  
mechanism will not work precisely for the minimal model  
(\ref{MGAmin}), as briefly discussed following equation (\ref{hpi}).  
Besides this, the divergence conditions (\ref{div}) are constraining  
enough that, when working in a {\it physical gauge}, they could rule out  
spacetimes of physical interest, including homogeneous, isotropic FRW  
cosmologies \cite{GregsTalk}. Some of these issues can already be seen  
in \cite{dRG3,new}. Thus, the classical solutions in massive gravity  
models with flat $\bar f_{\mu\nu}$ may drastically differ from general  
relativity.   
   
Of course, the action (\ref{MGAmin}) can also be used for non-flat   
$\bar f$, although in this case, the ghost problem has to be    
investigated afresh. Let us split the metric into {\it background} and    
{\it fluctuation} parts, $g_{\mu\nu}=\hat g_{\mu\nu}+h_{\mu\nu}$ and   
choose the physical gauge, $\phi^\mu=x^\mu$. Now take,    
\be   
f_{\mu\nu}=c^2\,\hat g_{\mu\nu}\,.   
\label{fg}   
\ee   
The caveat here is that, classically, there is no real distinction  
between the ``background'' and the ``fluctuation'' (unless  
$T_{\mu\nu}$ has a natural split in this way). In principle, this  
problem can be avoided by giving appropriate dynamics to $f_{\mu\nu}$.  
We will not attempt that here, except to say that in that case the  
$\sqrt{-\det f}$ term can no longer be disregarded. If we simply add a  
term $M^2_pm^2\lambda_f\sqrt{-\det f}$ to the action, then the  
$f_{\mu\nu}$ minimization condition,  
\be  
(\sqrt{g^{-1}f})f^{-1}+f^{-1}(\sqrt{g^{-1}f})^T-2\lambda_f\sqrt{\det  
  g^{-1}f}f^{-1}=0   
\ee   
has the solution $c=(1/\lambda_f)^{\frac{1}{3}}$. More generally, the  
above ansatz naturally arises as a solution in bi-gravity theories  
\cite{Blas,Afshar}. Here, we will not attempt giving $f_{\mu\nu}$ any  
specific dynamics, but will bear this picture in mind in what follows.  
The intention is to discuss the properties of solutions that resemble  
(\ref{fg}) in the appropriate extension of the theory.  
   
With the above ansatz, the divergence condition (\ref{div}) gives  
$\p_\mu c=0$ and the metric equation of motion (\ref{eom-g}) reduces   
to the massless Einstein's equation for $\hat g$ with an observable  
cosmological constant,      
\be   
\Lambda_{obs}=\Lambda-6m^2(1-c)\,.   
\ee    
Hence all background solutions are the same as in Einstein-Hilbert  
gravity with $\Lambda_{obs}$. However, metric fluctuations around this  
background are massive, with a mass term,     
\be   
\frac{1}{4}m_h^2\,\sqrt{-\hat g}\left(h^\mu_\nu h^\nu_\mu-(h^\mu_\mu)^2   
\right)\,, \qquad\qquad m_h^2=m^2c\,.   
\ee    
The indices are raised with $\hat g$. This can be read off from the  
minimal massive action after taking into account $\Lambda_{obs}$,  
which is itself read off from the metric equation (not from the  
action).      
   
Since $c>0$, tuning it can neutralize a positive $\Lambda\leq 6m^2$,   
as well as any negative $\Lambda$. Note that $m^2$ need not be small   
to recover the GR solutions and that the fluctuations are light as   
long as $c$ is small. In the extreme case, $m^2$ could be of the same   
order as a quantum field theory $\Lambda$ (say, at TeV scale) and   
$c<<1$ to cancel a large cosmological constant with effectively light   
gravitons. In the special case of $\Lambda=6m^2$, one ends up with a   
graviton mass of the same order as the observed cosmological constant,   
$\Lambda_{obs}\sim 6m_h^2$. The ability to cancel a $\Lambda$ in this   
way generalizes the mechanism of screening of cosmological constant,   
discussed for a flat spacetime in \cite{DHK,dRG3,HHvS,dR}, to any background   
sourced by any $T_{\mu\nu}$. What is more, since the classical   
solutions are the same as in GR, the light bending discrepancy between   
massless and massive gravity is automatically avoided for any source,   
without invoking the Vainshtein mechanism (although this mechanism may   
still apply to the fluctuations).   
   
In \cite{GS,BDH} the authors considered massive graviton excitations   
over an FRW solution of massless general relativity. They studied the   
ghost related stability bounds on the Hubble parameter in terms of the   
graviton mass, generalizing earlier work on massive gravity in de   
Sitter spacetime \cite{Higuchi}. It is easy to see that the ansatz   
(\ref{fg}) with $\hat g$ an FRW metric, provides precisely the   
non-linear framework in which the set up of \cite{GS,BDH} is   
realized. Of course, for the reasons mentioned above, to get a   
consistent picture, one has to include appropriate dynamics for   
$f_{\mu\nu}$. Then the theory should provide a consistent description   
beyond the regime of classical stability breakdown \cite{BDH}.   
   
A purpose of this discussion is to emphasize that the relevance of
vDVZ discontinuity \cite{vDVZ1,vDVZ2} to the observational viability
of massive gravity depends on the choice of $f_{\mu\nu}$. In general,
if this tensor is close enough to $g_{\mu\nu}$ for a certain classical
solution of the massive theory, this $g_{\mu\nu}$ in turn can be very
close to the corresponding solution in the massless theory. Thus the
observational constraints imposed by the vDVZ discontinuity can be
circumvented without invoking the Vainshtein mechanism\footnote{
  Although the massive fluctuations in such backgrounds may still
  exhibit the vDVZ discontinuity, they will not affect light bending.
  An exception is massive gravity in AdS backgrounds which is known to
  not exhibit the vDVZ discontinuity at all \cite{Porrati,KMP}.}. This
also underlines the need for further investigating the dynamics of
$f_{\mu\nu}$ which is beyond the scope of this paper.

\section{The General non-Linear Massive Action}   
   
In this section we show that the minimal massive action of the  
previous section can be generalized to, at most, a two-parameter  
family of non-linear actions. The generalizations are constructed in  
terms of a deformed determinant and correspond to adding interactions  
to the minimal theory. We elaborate on the formal structure of these  
terms. The generalized actions are ghost free at least to lowest  
orders in an expansion around flat space and for a flat auxiliary  
metric $f_{\mu\nu}$.  Hence any ghost free theory of massive gravity  
must belong to this class.  
   
\subsection{The generalized mass term as a deformed determinant}   
  
The massive action (\ref{MGA}) is the minimal deformation of  
$\det\sqrt{g^{-1}f}$ that satisfies the two criteria listed in the  
beginning of section $3$: (i) it is free of ghosts at the lowest order  
in the fields when perturbing around flat space and (ii) it reduces to  
the covariant Fierz-Pauli mass term at lowest order in $H^\mu_\nu$.  
Let us now consider the most general deformation of the determinant  
with deformation parameters $\alpha_n$.  Taking $\sqrt{g^{-1}f}=\mK  
+\one$ we define,\footnote{For the curved space $\epsilon$-tensors, we  
  use the conventions of Wald in {\it General Relativity} \cite{Wald},  
  page 432.}   
\be    
\widehat{\det}\sqrt{g^{-1}f}=\widehat{\det}(\mathbb 1 +\mK)=   
\sum_{n=0}^{4}\frac{-\alpha_n}{n ! (4-n)!}\,   
\epsilon_{\mu_1\cdots\mu_n\lambda_{n+1}\cdots\lambda_4}\,   
\epsilon^{\nu_1\cdots\nu_n\lambda_{n+1}\cdots\lambda_4}\,   
\mK^{\mu_1}_{\;\;\nu_1}\cdots \mK^{\mu_n}_{\;\;\nu_n}\,.   
\label{Defdet}    
\ee   
The ordinary determinant corresponds to $\alpha_n=1$.  For future  
convenience we note that the deformed determinant (\ref{Defdet}) can  
be rearranged to give an equivalent expression in terms of  
$\sqrt{g^{-1}f}$, rather than $\mK$,  
\be    
\widehat{\det}\sqrt{g^{-1}f}=\!\!   
\sum_{r=0}^{4}   
\frac{-\beta_r}{r!(4-r)!}   
\epsilon_{\mu_1\cdots\mu_r\lambda_{r+1}\cdots\lambda_4}   
\epsilon^{\nu_1\cdots\nu_r\lambda_{r+1}\cdots\lambda_4}   
(\sqrt{g^{-1}f})^{\mu_1}_{\,\nu_1}\cdots   
(\sqrt{g^{-1} f})\,^{\mu_r}_{\,\nu_r}\,,   
\label{MassDD}    
\ee   
where,   
\be   
\label{beta}   
\beta_r = (4-r)!\sum_{n=r}^4\,\frac{(-1)^{n+r}}{(4-n)!(n-r)!}   
\,\alpha_n\,.   
\ee   
It will often be easier to manipulate the deformed determinant in this form.   
   
The claim is that, in terms of the deformed determinant, the full  
non-linear action is given by,    
\be   
S=-M_p^2\int d^4x \sqrt{-g}\left[     
R -2m^2\,\widehat{\det}(\sqrt{g^{-1}f})+\Lambda'\right]\,.   
\label{MGAmax}   
\ee   
This action automatically satisfies the two criteria specified above:  
At the lowest order in the metric perturbation around flat space, and  
for $\bar f_{\mu\nu}= \eta_{\mu\nu}$, we have $\mK^\mu_{\;\;\nu}  
\big\vert_{h=0}=\p^\mu\p_\nu\pi$.  Then (\ref{Defdet}) reduces to the  
most general total derivative term for $\pi$, insuring the absence of  
ghosts at this order.  Also, the generalized mass term contains the  
minimal mass term (\ref{MGA}) for the right set of parameters and  
hence reduces to the covariant Fierz-Pauli action (\ref{FPCov}).  
   
The five coefficients $\alpha_n$ (or equivalently $\beta_n$) in fact  
represent only two free parameters, besides the graviton mass $m$ and  
the cosmological constant $\Lambda$.  The lowest order term $\alpha_0$  
can be absorbed into a cosmological constant as was done in  
(\ref{MGAmin}).  For clarity, we will define $\alpha_0$ in such a way that  
flat spacetime can be considered a valid background when the  
cosmological constant $\Lambda$ is zero.  This means that the mass  
term given by the deformed determinant should contain no terms linear  
in the metric fluctuation $h$.  This is equivalent to setting  
$\alpha_0 = \alpha_1$.  Then the total effective cosmological constant  
is given by $\Lambda$.  
   
A second coefficient can be fixed by the appropriate definition of the
graviton mass $m^2$.  That is, we define the $\alpha_n$ such that the
term quadratic in $h_{\mu\nu}$ becomes the FP mass term (\ref{FP}).
This is equivalent to setting $\alpha_2-\alpha_0 = 1$.  A third
coefficient can be eliminated by observing that since the undeformed
determinant $\det \sqrt{g^{-1} f}$ does not contribute to the
equations of motion it can be used to subtract off one term of the
deformed expansion.  We will show this below.  Thus the generalized
action depends on only two free parameters due to its anti-symmetric
structure. The deformed determinant does not allow for any further
generalization. The recursion relation constructed in \cite{dRGT}, and
the vanishing of the higher order terms arising from it, can be
understood in terms of this structure, as will be explained below.
   
The proof of the absence of ghosts to quartic order in non-linearities  
reduces to the calculation given in \cite{dRGT} and will not be  
repeated here.  The presence of the square-root matrix complicates the  
analysis of the general case, but there are indications that the ghost  
analysis can be extended to higher orders in an ADM formalism.  The  
general action in the form (\ref{MGAmax}) also allows one to go beyond  
the flat auxiliary metric and consider classical solutions for more  
general $f_{\mu\nu}$. This allows for a richer class of solutions,  
however, now the absence of ghost has to be checked separately for  
different choices of $f_{\mu\nu}$. The structure of the general  
massive action will be discussed in more detail below.  
  
\subsection{Formal construction}   
  
That (\ref{MGAmax}) is the most general massive gravity action needs a 
little more clarification. The deformed determinant contains terms 
that are at most fourth order in $\sqrt{g^{-1} f}$. It is natural to 
ask whether the Lagrangian can contain higher order terms. These terms 
would have to vanish identically when evaluated at $h_{\mu \nu} =0$ in 
order to be consistent with the ghost-free construction. The epsilon 
tensor structure of the deformed determinant above would suggest that 
no such terms exist. However, in \cite{dRGT} a recursion relation is 
presented that might seem to allow one to construct higher order terms 
in the Lagrangian. In what follows we will show explicitly that these higher 
order terms are identically zero in 4 dimensions.  
 
The vanishing of these terms can be simply understood in the following 
way: Writing $\det(\mathbb 1+\mX)$ as $e^{\Tr\ln(\mathbb 1+\mX)}$ for 
a general matrix $\mX$, and expanding the exponential in powers of 
$\mX$ gives an infinite number of terms. For a $4\times 4$ matrix, 
only terms to order $4$ are non-zero by the definition of determinant. 
These terms appear in the construction (\ref{Defdet}). The higher 
order terms would contribute in higher dimensions, but vanish for 
$d=4$. These are the terms constructed by the recursion relation given 
below in (\ref{Newt}). 
    
We start by considering a general $N \times N$ matrix $\mX$.  The  
determinant of $\mathbb 1+\mX$ can be written in terms of the  
eigenvalues $\lambda_i$ of the matrix $\mX$,  
\be   
\det (\mathbb 1+\mX) = \prod_{i=1}^N (1+\lambda_i)   
=\sum_{k=0}^N  e_k(\lambda_1, \ldots, \lambda_N) \, .   
\label{X}  
\ee   
The terms $e_k(\lambda_1, \ldots, \lambda_N)$ are elementary symmetric  
polynomials of the eigenvalues.  That is, they are the sum of all  
distinct products of $k$ distinct $\lambda_i$,  
\be   
\begin{array}{lcl}   
e_0(\lambda_1, \ldots, \lambda_N)&=&1 , \\   
e_1(\lambda_1, \ldots, \lambda_N)&=&\lambda_1+\ldots+\lambda_N \, , \\   
e_2(\lambda_1, \ldots, \lambda_N)&=& \sum_{i<j} \lambda_i \lambda_j , \\   
\vdots && \\   
e_N(\lambda_1, \ldots, \lambda_N)&=&\lambda_1\lambda_2 \ldots  
\lambda_N = \det \mX \, , \\    
e_k(\lambda_1, \ldots, \lambda_N)&=&0 ~~{\rm for}~ k>N \, . \\   
\end{array}   
\ee   
The eigenvalues can also be used to represent the trace of powers of  
the matrix $\mX^m$.  Here we will use square brackets to denote the  
trace,  
\be   
[\mX^m] = \sum_{i=1}^N \lambda_i^m  \, .   
\ee   
Newton's identities give a relationship between the $e_k(\lambda_1,  
\ldots, \lambda_N)$ and the trace terms,    
\be   
e_k(\lambda_1, \ldots, \lambda_N)= - \frac{1}{k} \sum_{m=1}^k (-1)^m  
[\mX^m] \, e_{k-m}(\lambda_1, \ldots, \lambda_N) \, .   
\label{Newt}   
\ee   
These identities can be used to write the polynomials $e_k(\lambda_1,  
\ldots, \lambda_N)$ entirely in terms of the traces $[\mX^m]$.  For a  
generic $4\times 4$ matrix one finds,  
\be   
\label{en}   
\begin{array}{lcl}   
e_0(\mX)&=&1  \, , \\   
e_1(\mX)&=&[\mX]  \, , \\   
e_2(\mX)&=&\tfrac{1}{2}([\mX]^2-[\mX^2]), \\   
e_3(\mX)&=&\tfrac{1}{6}([\mX]^3-3[\mX][\mX^2]+2[\mX^3]) \, , \\   
e_4(\mX)&=&\tfrac{1}{24}([\mX]^4-6[\mX]^2[\mX^2]+3[\mX^2]^2   
+8[\mX][\mX^3]-6[\mX^4])\, , \\   
e_k(\mX)&=&0 ~~{\rm for}~ k>4 \, . \\   
\end{array}   
\ee   
The $e_k(\mX)$ are exactly the terms that appear in the deformed  
determinant (\ref{Defdet}) as can be seen from (\ref{X}). Replacing  
$\mX$ by $\mK=\sqrt{g^{-1} f}-\one$ and setting $N=4$ we can write,  
\be   
\widehat{\det}\sqrt{g^{-1} f}=   \sum_{n=0}^{4} \alpha_n e_n(\mK) \, .   
\ee   
Given this formulation, it is apparent that the recursion relation 
given in \cite{dRGT} is equivalent to Newton's identities (\ref{Newt}) 
with the identification ${\cal L}^{(n)}_{der} \rightarrow n! \, 
e_n(\mK)$. For any $N \times N$ matrix this same recursion relation 
gives $0$ for $k>N$. Thus the higher order terms in $\sqrt{g^{-1} f}$ 
that give the correct ghost free structure in the decoupling limit 
must be zero for $n>4$. 
   
The existence of only two free parameters can also be seen at the  
level of the perturbative Lagrangian presented in \cite{dRG2}. The  
first two parameters of that theory can be related to the $\alpha_n$  
in the deformed determinant as $c_3= \frac{1}{6}(\alpha_3-\alpha_0)$  
and $d_5=-\tfrac{1}{48}(\alpha_4-\alpha_0)$. However, using the above  
results, the terms multiplying $f_7$ can be shown to vanish. To see 
this, note that $e_5(\mX) =0 $ for a generic $4 \times 4$ matrix. 
Then, from Newton's identities, the following combination of terms 
must be identically zero,   
\be   
[\mX]^5-10[\mX]^3[\mX^2]+15[\mX][\mX^2]^2-20[\mX^2][\mX^3]+  
20[\mX]^2[\mX^3]-30[\mX][\mX^4]+24[\mX^5] = 0 \, .   
\ee   
If we take $\mX$ to be the matrix $\mH$, then the coefficient $f_7$ in  
\cite{dRG2} appears in the action multiplying precisely this  
combination of terms. Analogous arguments can be used to eliminate  
free parameters that appear at higher orders in $\mH$. Thus in four  
dimensions there are only two free physically meaningful parameters in  
the massive action. The higher order terms in an $\mH$ expansion can  
be determined entirely from the coefficients $c_3$ and $d_5$.  
   
\subsection{Equivalent descriptions of the action}   
  
The full non-linear action (\ref{MGAmax}) can be written in terms of  
the $e_n(\mK)$ given by (\ref{en}) above:  
\be   
\label{actnow}   
S=-M_p^2\int d^4x\sqrt{-g}\,R(g) + 2 M_p^2 m^2\int d^4x\sqrt{-g}\,   
\sum_{n=0}^{4} \alpha_n e_n(\mK) \, .   
\ee   
With this formalism, it is straight-forward to convert between  
equivalent descriptions of the action. Let us first make contact with  
the Lagrangians presented in \cite{dRG1,dRG2,dRGT}.  
   
To eliminate the redundant parameters we rewrite,   
\be   
\alpha_0 e_0(\mK) = \alpha_0 \det \sqrt{g^{-1} f} -\sum_{n=1}^{4}   
\alpha_0 e_n(\mK)\, .   
\ee   
The potential term of the action becomes,   
\be   
2 M_p^2 m^2\int d^4x\sqrt{-g}\, \left\{\sum_{n=1}^{4} \bar{\alpha}_n e_n(\mK)    
+\alpha_0 \det \sqrt{g^{-1} f} \right\}\, ,   
\ee   
where $\bar{\alpha}_n \equiv \alpha_n-\alpha_0$. As shown above, the  
determinant term $\alpha_0 \det (\sqrt{g^{-1} f})$ is non-dynamical  
and can be dropped from the action. In order for flat spacetime to be  
a valid solution, terms linear in $h$ must vanish in the action when  
expanded around $\eta_{\mu \nu}$. It follows that $\bar{\alpha}_1 =  
0$. Finally, fixing the coefficient of the mass term to be of the  
canonical form gives $\bar{\alpha}_2 =1$. The action (\ref{actnow})  
reduces to  
\be   
\label{actlater}   
S=-M_p^2\int d^4x\sqrt{-g}\,R(g) + 2 M_p^2 m^2\int  
d^4x\sqrt{-g}\,(e_2(\mK)+ \bar{\alpha}_3 e_3(\mK)+\bar{\alpha}_4 e_4(\mK))\, .   
\ee   
This action has two free parameters, $\bar{\alpha}_3$ and  
$\bar{\alpha}_4$. Written in this way it is easy to see that the  
action recovers the covariant Fierz-Pauli form when expanded in $\mH$.  
In this form it can be compared to the results of  
\cite{dRG1,dRG2,dRGT}. When the parameters $\bar{\alpha}_3$ and  
$\bar{\alpha}_4$ are set to zero, one obtains the minimal resummed  
theory presented in \cite{dRGT}. When $\bar{\alpha}_3 = \bar{\alpha}_4  
=1$ one obtains the minimal deformed determinant action presented  
above. More generally, the parameters $\bar{\alpha}_3$ and  
$\bar{\alpha}_4$ are related to the coefficients $c_3$ and $d_5$ of  
\cite{dRG2} by $ \bar{\alpha}_3 = 6 c_3$ and $ \bar{\alpha}_4 = -48  
d_5$.  
   
Because of the non-dynamical nature of the $\det \sqrt{g^{-1} f}$  
term, the action (\ref{actlater}) has an equivalent description in  
terms of the lower order $e_n$. Moreover, we can use the definitions  
(\ref{MassDD}) and (\ref{beta}) to rewrite the action in terms of  
$e_n(\sqrt{g^{-1} f})$ rather than $e_n(\mK)$. When manipulating the  
action non-linearly it will be more convenient to work with this
simpler description. After some rearranging, the action is given by   
\be   
\label{actlater2}   
S=-M_p^2\int d^4x\sqrt{-g}\,R(g) +2 M_p^2 m^2\int d^4x\sqrt{-g}\,   
\sum_{n=0}^{3} \beta_n\, e_n(\sqrt{g^{-1} f})\, ,   
\ee   
where the $\beta_n$ are given by (\ref{beta}):   
\bea   
\beta_0& =& 6-4\bar{\alpha}_3+\bar{\alpha}_4 \, ,\nonumber \\   
\beta_1 &= &-3+3\bar{\alpha}_3-\bar{\alpha}_4 \, ,\nonumber \\   
\beta_2 &= &1-2 \bar{\alpha}_3+\bar{\alpha}_4 \, ,\nonumber \\   
\beta_3 &=& \bar{\alpha}_3-\bar{\alpha}_4\, .   
\eea   
This action contains terms that are at most third order in  
$\sqrt{g^{-1} f}$ rather than fourth order as in (\ref{actlater}).  
When expanded in $h_{\mu\nu}$ this action reduces to the basic  
form of the Fierz-Pauli mass (\ref{FP}) at lowest order. The minimal  
theory presented in the previous section corresponds to the choice of  
coefficients $\beta_0 = 3$, $\beta_1 = -1$ and $\beta_2 = \beta_3 = 0$  
(i.e., $\bar{\alpha}_3 = \bar{\alpha}_4 =1$). We stress that even  
though this action does not reduce to the {\it covariant} Fierz-Pauli  
form, the difference is cosmetic and the actions (\ref{actlater}) and  
(\ref{actlater2}) are physically equivalent (as long as $\bar  
f_{\mu\nu}$ is non-dynamical).   
   
For completeness we also present here the full non-linear equations of  
motion.  To vary the action we use the relationship,  
\be   
\delta\Tr\left[\left(\sqrt{g^{-1} f}\right)^n\right] = \frac{n}{2} \,   
\Tr\left[g \left(\sqrt{g^{-1} f}\right)^n \delta g^{-1}\right] \, .   
\ee   
Then it follows that,   
\be   
\frac{2}{\sqrt{-g}}\delta\left(\sqrt{-g}\,e_n(\sqrt{g^{-1}f})\right)    
= \sum_{m=0}^n (-1)^{m+1}\Tr\left[g \left(\sqrt{g^{-1} f}\right)^m   
\delta g^{-1}\right]e_{n-m}(\sqrt{g^{-1} f})\, .   
\ee   
Then the variation of the action gives,   
\be   
R_{\mu \nu}-\frac{1}{2}  g_{\mu \nu} R+\frac{m^2}{2} \sum_{n=0}^{3} (-1)^n  
\,\beta_n\,\left[g_{\mu \lambda}\, Y_{(n)\nu}^{\,\,\,\lambda}+g_{\nu\lambda} 
\, Y_{(n)\mu}^{\,\,\,\lambda} \right] =G_N T_{\mu \nu}\, ,   
\ee   
where $Y_{(n)} \equiv Y_{(n)} (\sqrt{g^{-1}f})$ and for $\mX
=\sqrt{g^{-1}f} $ we have defined (with square brackets again denoting
the trace),   
\bea   
Y_{(0)}(\mX) &\equiv& \one \, ,\nonumber \\[.1cm]   
Y_{(1)}(\mX) &\equiv& \mX-\one[\mX] \, , \\[.1cm]      
Y_{(2)}(\mX)&\equiv& \mX^2-\mX  
    [\mX]+\tfrac{1}{2}\one ([\mX]^2-[\mX^2]) \, ,\nonumber \\[.1cm]      
Y_{(3)}(\mX) &\equiv& \mX^3-\mX^2[\mX]+\tfrac{1}{2}   
\mX ([\mX]^2-[\mX^2])   
 -\tfrac{1}{6}\one([\mX]^3-3[\mX][\mX^2]+2[\mX^3]) \,.\nn 
\eea   
  
\section{Discussion of results and conclusions}   
   
Our main results are summarized in section 1.  Here we discuss   
some of their implications, and related issues.   
   
{\bf How general is the setup?} One might wonder about the generality 
of the non-linear massive gravity actions constructed via the methods 
of \cite{AGS,CNPT,dRG2,dRGT}. We have argued in this work that in fact 
these methods are more powerful than they may initially appear, in 
more than one way. The actions in \cite{AGS,CNPT,dRG2,dRGT} are 
constructed so that they reduce to the covariant FP form (\ref{FPCov}) 
at the lowest order (quadratic in this case) in $\mH$. At first sight, 
they seem to exclude mass terms that reduce to (\ref{MGAminH}), even 
though these are completely consistent with the basic FP mass 
(\ref{FP}). However, we have shown that the excluded actions differ 
from the rest only by non-dynamical terms and, in this sense, are not 
left out. Also, within our formulation the origin of the 2-parameter 
family of actions, and the impossibility of adding more parameters, 
becomes manifest. 
 
Moreover, mass terms considered in \cite{AGS,CNPT,dRGT} and here,  
depend only on scalar functions of the form $F(g^{-1}f)$. One may  
wonder about mass terms that involve scalar functions $F'(f^{-1}g)$.  
As pointed out in section 2, the quintic order ghost analysis of  
\cite{dRG2} is general enough to allow for both forms. However, it so  
happens that at the non-linear level, $F(g^{-1}f)$ is adequate to  
express all admissible (by the no-ghost condition) non-linear mass  
terms in a simple way.   
   
Of course, the theory presented here can be extended by introducing   
dynamics for $f$, considering additional auxiliary metrics or other   
such modifications. However, any theory of massive gravity that  
reduces to the basic FP action at lowest order in $h$, is  
ghost-free to lowest order in interaction, and depends only on the  
metric $g$ and the auxiliary metric $f$, must belong to the  
two-parameter family of actions presented here.   
   
{\bf Non-flat auxiliary metric}. The ghost analysis of massive gravity  
actions is valid only for a flat auxiliary metric.  The associated  
constraints introduce deviations from general relativity. For such  
theory to comply with observations, at the least one needs the  
Vainshtein mechanism \cite{V,DDGV}. On the other hand, if one allows  
the auxiliary metric $f_{\mu\nu}$ to be non-flat, or better, related  
to $g_{\mu\nu}$ as in section 3.3, then the classical solutions could  
be very close to general relativity while still retaining some  
attractive features of massive gravity, like the screening of the  
cosmological constant. The proper arena of realizing such solutions is  
theories with dynamical $f_{\mu\nu}$, though not necessarily  
restricted to bi-gravity theories. The ghost issue in such models has  
to be re-examined. In fact, the recent analysis of massive gravity in  
FRW spacetimes \cite{GS,BDH} is relevant to this construction.  
   
{\bf The ghost issue.} It is clear that the appearance of the   
square-root matrix $\sqrt{g^{-1}f}$ is an inevitable feature of the   
Goldstone boson formulation of massive gravity \cite{AGS}. Such a   
matrix cannot be easily manipulated analytically. Thus the very setup   
that simplifies the ghost analysis in the decoupling limit, introduces   
a complication at the non-linear level. A straightforward ADM analysis   
of the ghost is therefore not an option, although such an analysis can   
be carried out for some special cases, for example, for $N_i=0$. For   
more on this, see \cite{dRGT}. (also see the note added below)  
   
Another issue is that even for models with flat $f_{\mu\nu}$, where  
the ghost analysis applies, the 2-parameter family of actions may need  
to be further constrained to a single parameter if one is to avoid  
unstable non-asymptotically flat solutions, as can be seen from  
\cite{dRG3}. Thus, while the candidate ghost-free actions considered  
in \cite{dRG2,dRGT} and here are a step forward, the final  
question of non-linear massive gravity is far from settled.   
 
\vspace{.3cm} 
\noindent {\bf Note added:}  After this work was submitted to the
arXiv, \cite{TMN} appeared with some related results. More recently
the absence of ghost at the non-linear level was shown in
\cite{HR3}.   
   
\acknowledgments   
   
We would like to thank Marcus Berg, Cedric Deffayet, Claudia de Rham,  
Denis Dietrich, Jonas Enander, Gregory Gabadadze, Stefan Hofmann,  
Eugene Lim, Edvard M\"ortsell, Stefan Sj\"ors, Angnis Schmidt-May, Bo  
Sundborg, Andrew Tolley and Mikael von Strauss for useful discussions.  
RAR is supported by the Swedish Research Council (VR) through the  
Oskar Klein Centre.

\end{document}